\documentclass[twocolumn,superscriptaddress,aps,amsmath,amssymb,prb]{revtex4}
\usepackage{bm}

\newcommand{\B}{\mbox{\tiny B}}
\newcommand{\T}{\mbox{\tiny T}}
\newcommand{\s}{\mbox{\tiny S}}
\newcommand{\D}{\mbox{\tiny D}}

\newcommand{\ti}{\Tilde}
\newcommand{\nl}{\nonumber \\}
\newcommand{\Sch}{Schr\"{o}dinger\ }

\newcommand{\Sec}[1]{Sec.\;\ref{#1}}
\newcommand{\App}[1]{Appendix\;\ref{#1}}
\newcommand{\be}{\begin{equation}}
\newcommand{\ee}{\end{equation}}
\newcommand{\bea}{\begin{eqnarray}}
\newcommand{\eea}{\end{eqnarray}}
\newcommand{\bsube}{\begin{subequations}}
\newcommand{\esube}{\end{subequations}}
\newcommand{\Eq}[1]{Eq.\,(\ref{#1})}
\newcommand{\Eqs}[1]{Eqs.\,(\ref{#1})}

\newcommand{\dg}{\dagger}
\newcommand{\la}{\langle}
\newcommand{\ra}{\rangle}

\newcommand{\ind}{{\sf n}}
\newcommand{\bn}{\bar n}

\newcommand{\bfalp}{\bm\alpha}

\newcommand{\rhonswap}{\rho_{\sf n}^{{ }_{\{\leftrightarrows\}}}}
\newcommand{\rhonup}{\rho_{\sf n}^{{ }_{\{+\}}}}
\newcommand{\rhondown}{\rho_{\sf n}^{{ }_{\{-\}}}}
\begin{document}

\title{Dynamics of quantum dissipation systems interacting
  with Fermion and Boson grand canonical bath ensembles:
   Hierarchical equations of motion approach}

\author{Jinshuang Jin}
\email[Corresponding authors; ]{jinjs@ust.hk, yyan@ust.hk}
\affiliation{Department of Chemistry, Hong Kong University
   of Science and Technology, Kowloon, Hong Kong}
\author{Sven Welack}
\affiliation{Department of Chemistry, Hong Kong University
   of Science and Technology, Kowloon, Hong Kong}
\author{JunYan Luo}
\affiliation{State Key Laboratory for Superlattices and Microstructures,
 Institute of Semiconductors, Chinese Academy of Sciences,
 P.O.\ Box 912, Beijing 100083, China}
\affiliation{Department of Chemistry, Hong Kong University
   of Science and Technology, Kowloon, Hong Kong}
\author{Xin-Qi Li}
\affiliation{State Key Laboratory for Superlattices and Microstructures,
 Institute of Semiconductors, Chinese Academy of Sciences,
 P.O.\ Box 912, Beijing 100083, China}
\affiliation{Department of Chemistry, Hong Kong University
   of Science and Technology, Kowloon, Hong Kong}
\author{Ping Cui}
\affiliation{Hefei National Laboratory for Physical Sciences
  at Microscale, University of Science and Technology
 of China, Hefei, China}
\affiliation{Department of Chemistry, Hong Kong University
   of Science and Technology, Kowloon, Hong Kong}
\author{Rui-Xue Xu}
\affiliation{Hefei National Laboratory for Physical Sciences
  at Microscale, University of Science and Technology
 of China, Hefei, China}
\affiliation{Department of Chemistry, Hong Kong University
   of Science and Technology, Kowloon, Hong Kong}
\author{YiJing Yan$^{\ast}$}
\affiliation{Hefei National Laboratory for Physical Sciences
  at Microscale, University of Science and Technology
 of China, Hefei, China}
\affiliation{Department of Chemistry, Hong Kong University
   of Science and Technology, Kowloon, Hong Kong}

\date{Submitted to JCP on 4 December 2006; revised on 2 February}

\begin{abstract}
 A hierarchical equations of motion formalism
 for a quantum dissipation system in a grand canonical bath
 ensemble surrounding is constructed, on the basis of the
 calculus-on-path-integral algorithm, together with the
 parametrization of arbitrary non-Markovin bath
 that satisfies fluctuation-dissipation theorem.
   The influence functionals for both the
 Fermion or Boson bath interaction are found to be of
 the same path-integral expression as the canonical bath, assuming
 they all satisfy the Gaussian statistics.
 However, the equation of motion formalism are different, due to
 the fluctuation-dissipation theories that are distinct and used
 explicitly.
 The implications of the present work to quantum
 transport through molecular wires and electron transfer
 in complex molecular systems are discussed.
\end{abstract}

\maketitle

\section{Introduction}
\label{thintro}

   Quantum dissipation theory (QDT) is concerned
 with the fundamental formulations
 for the dynamics of a quantum system of
 primary interest embedded in a quantum bath
 surrounding environment.
 The key quantity in QDT is the so-called reduced system
 density operator,
  $\rho(t) \equiv {\rm tr}_{\B}\rho_{\rm T}(t)$,
 i.e., the partial trace of the
 total density operator over the
 bath space of practically infinite degrees of freedom.
 Due to its fundamental importance and intrinsic complexity,
 the development of QDT has involved scientists from
 many fields of research,%
 \cite{Kub85,Muk95,Wei99,Wal94,Dit98,Ors00,Bre02,Nit06,Kle06,Han05026105}
  but it has remained as a
  challenging topic since the middle of the last century.

   The exact $\rho(t)$ can be expressed in terms of the Feynman-Vernon
 influence functional, assuming the interaction bath
 satisfies the Gaussian statistics.\cite{Fey63118}
 Its numerical implementation has been carried out in a few simple systems
 with the forward--backward iterative path integral
 propagation method.\cite{Mak952430,Mak984414}
 Unraveling the path integral formalism into
 stochastic \Sch or stochastic Liouville equations
 has also been proposed  in various forms
 in the past 10 years.%
 \cite{Kle95224,Dio97569,Ple98101,Str991801,Bre991633,Bre04022115,%
 Sto994983,Sto01249,Sto02170407,Sha045053,Yan04216,Sha06187}
 The main obstacle to both the path integral and
 stochastic differential equation QDTs is their formidable numerical
 implementation in multilevel systems.
 Applications based on the path integral or stochastic
 QDT formalism are also often too cumbersome to be practical.

   For the numerical efficiency and practical use in general,
 one will be interested in a differential equations of motion
 (EOM) formalism. In principle, a formally exact (but unclosed)
 EOM formalism of QDT can be constructed via the
 Nakajima-Zwanzig-Mori projection
 operator technique.\cite{Nak58948,Zwa61,Mor65423,Arg64A98}
 To complete the formalism, however, one usually has to invoke
 certain approximations. In particular, various commonly used
 forms of Bloch-Redfield theory and Fokker-Planck equations%
 \cite{Red651,Gra88115,Cal83374,Leg871,Yan002068,Xu037,Yan05187}
 involve the Markovian limit and the weak system-bath interaction.
 The validities of these approximations are increasingly challenged,
 especially due to the emerging fields of nanoscience
 such as quantum transport and quantum information
  processes.\cite{Gur9615932,Gur9715215,Gur986602,Li04085315,%
 Li05066803,Li05205304,Jin05143504,Sta04136802,%
 Goa01125326,Shn9815400,Moz02161313,Tom03256801,Wel06044712}

 We have recently constructed a formally exact
 EOM formalism for arbitrary non-Markovian dissipation systems
 interacting with Gaussian canonical bath ensemble.\cite{Xu05041103,Xu07PRE}
 This theory generalizes the Tanimura and coworkers'
 hierarchical formalism.\cite{Tan89101,Tan914131,Ish053131,Tan06082001}
  The present paper is to extend our previous results
 to include both the Fermion and Boson grand canonical bath
 ensemble cases. The desired hierarchical EOM
 will be constructed, via the auxiliary
 influence-generating functionals (IGF) approach,
 together with the calculus-on-path-integrals
 (COPI) algorithm.\cite{Xu05041103,Xu07PRE}

   The remainder of this paper is organized as follows.
 In \Sec{thmodel}, after the general
 description of a system--bath
 composite Hamiltonians, we discuss
 the fluctuation-dissipation theorem
 and other useful quantum mechanics
 relations for the grand canonical ensembles
 of Fermion/Boson particles.
 In \Sec{thpath}, we revisit the influence functionals
 path--integral formalism of QDT,
 assuming that the interaction grand canonical bath
 satisfies Gaussian statistics.
 The derivations in relation to
 these two sections
 are detailed in \App{thapp_FDT}
 and \ref{thapp_path}, respectively.
 In \Sec{thparameter}, we present
 a bath correlation function
 parametrization scheme
 that will be used for the
 later development of differential
 EOM formalism for general non-Markovian
 dissipation systems.
  In \Sec{thspecial}, we exploit a
 model dissipation system
 to illustrate the key ingredients
 in the IGF-COPI construction of the
 hierarchical EOM, augmented with
 residue correction and truncation.
 The final exact EOM formalism for general cases
 is presented in \Sec{thgeneral}.
  It is in principle applicable for
 arbitrary non-Markovian dissipation systems,
 interacting with the Fermionic/Bosonic
 grand canonical bath ensemble
 at any temperature, including $T=0$.
 Finally, \Sec{thsum} presents the comments,
 discussions, and concluding remarks of this work.

 \section{Total composite Hamiltonian and
    fluctuation-dissipation theory}
 \label{thmodel}

\subsection{The description of system--bath coupling}
\label{thmodelA}

  Let us start with the
 total system--plus--bath composite Hamiltonian,
 which in the stochastic bath $h_{\B}$-interaction picture
 is given by
 \be \label{HT0}
    H_{\rm T}(t) = H(t) +\sum_a
  \left[W_{a} \hat f^{\dg}_{a}(t)+ \hat f_{a}(t)W^{\dg}_{a} \right] .
 \ee
 The deterministic Hamiltonian $H(t)$ for the reduced
 system may also involve a time-dependent external
  coherent field drive.
 The second term  in the right--hand--side (rhs)
 of \Eq{HT0} denotes the
 stochastic system-bath interaction.
 It is expressed in the multiple dissipative mode
 decomposition form, where $\{W_{a}\}$ and
 $\{f_{a}(t)=e^{ih_{\B}t}f_{a}e^{-ih_{\B}t}\} $
 are the system and bath operators, respectively.
 They are in general non-Hermitian.

   The interaction bath operators $\{f_a(t)\}$
 are also assumed to the Gaussian stochastic
 processes ({\it a la} the central limit theorem in statistics).
 The required Gaussian statistics is satisfied strictly
 when the individual $f_a$ ($f_a^{\dg}$) is a
 linear combination of the annihilation (creation) operators of
 the uncorrelated Fermion/Boson bath particles.
 The nonperturbative QDT formalisms to be
 developed later in this work will involve no further
 approximation.

 Throughout this work, we set $\hbar \equiv 1$ and the inverse
 temperature $\beta \equiv 1/(k_{\B}T)$. Denote
 $\la \hat O \ra_{\B} \equiv {\rm tr}_{\B}(\hat O\rho^{\rm eq}_{\B})$,
 with the grand canonical equilibrium density operator of the
 bare bath being given by
 \be\label{rhoeq}
   \rho^{\rm eq}_{\B}=\frac{ e^{-\beta (h_{\B}-\mu \hat N)}}
   {{\rm tr}_{\B}[e^{-\beta (h_{\B}-\mu \hat N)}]}.
 \ee
 Here, $\mu$ denotes the chemical potential of the bath.
 The particle number operator $\hat N=\sum_k c^\dg_kc_k$
 of the bath reservoirs commutes with the bare bath
 Hamiltonian, $[\hat N,h_{\B}]=0$.
  In the case of a canonical bath ensemble,
 the number of particles is
 conserved and no longer a dynamical variable. The resulting
  canonical $\rho^{\rm eq}_{\B}\propto e^{-\beta h_{\B}}$ is
 equivalent to set the bath chemical potential $\mu=0$.

  For the later development, we also recast the system-bath
 coupling, i.e.\ the second term in the rhs of \Eq{HT0}, as
 \be \label{Hprmt}
  H'(t)=\sum_{a,\sigma}W^{\bar\sigma}_a\hat f^{\sigma}_a(t).
 \ee
 Here, $W^{\bar\sigma}_a\equiv (W^{\sigma}_a)^\dg$
 that can be either $W_a$ or $W^\dg_a$;
 and similarly for
 $\hat f^{\bar\sigma}\equiv(\hat f^{\sigma})^{\dg}$.
 To be more specific, the bath operators
 $\{\hat f_a\}$ considered in this work
 are linear combinations of the annihilation operators,
 $\hat f_a=\sum_k t_{ak}c_k$, with
  $c_jc^{+}_k\pm c^{+}_kc_j = \delta_{jk}$
 for the Fermion ($+$) or Boson ($-$) particles in the
 grand canonical bath reservoirs.

 Note that as the stochastic bath operators $\hat f^{\bar \sigma}_a(t)
  =[\hat f^{\sigma}_a(t)]^{\dg}$ are the
 linear combinations of either the creation or the annihilation operators
 of the particles, we have
  $\la  \hat f^{\sigma}_{a}(t) \ra_{\B}
  = \la  \hat f^{\sigma}_{a}(t)\hat f^{\sigma}_{b}(\tau) \ra_{\B}=0$.
 These stochastic bath interaction operators
 are also assumed to follow the Wick's theorem for
 the thermodynamic Gaussian average that depends only
 on the two-time correlation functions,
\bsube\label{ffcorr0}
\bea
 C^{(+)}_{ab}(t-\tau) \!\!&=&\!\!
  \la\hat f^{\dg}_{a}(t)\hat f_{b}(\tau) \ra_{\B},
 \\
 C^{(-)}_{ab}(t-\tau) \!\!&=&\!\!
   \la\hat f_{a}(t)\hat f^{\dg}_{b}(\tau)\ra_{\B}.
\eea
\esube

\subsection{Fermion versus Boson bath reservoirs:
   Fluctuation-dissipation theorem}
\label{thmodelB}
   Now we present
 the fluctuation-dissipation theorem (FDT),
 and the related symmetry and detailed-balance relations
 for the involving interaction bath correlation functions.
 The key steps to the final results will be detailed in \App{thapp_FDT}.
 To avoid the confusion of later using
 $\pm$ to represent Fermion/Boson bath ensemble,
 let us recast \Eqs{ffcorr0} as
 \bea \label{ffcorr}
   C^{(\sigma)}_{ab}(t)
    = \la  \hat f^{\sigma}_{a}(t)  \hat f^{\bar\sigma}_{b}(0) \ra_{\B}.
 \eea
 They satisfy the symmetry and the detailed-balance
 relations as [cf.\ \Eqs{appFDT_sym}--(\ref{appFDT_detbal})]
 \be\label{Ctsym}
   [C^{(\sigma)}_{ ab}(t)]^\ast = C^{(\sigma)}_{ ba}(-t)
    = e^{\sigma\beta\mu} C^{(\bar\sigma)}_{ab}(t-i\beta).
 \ee
 Their frequency-domain counterparts are
 \be\label{Cwsym}
   C^{(\sigma)}_{ab}(\omega) = [C_{ba}^{(\sigma)}(\omega)]^\ast
  =e^{\beta(\omega+\sigma\mu)} C^{(\bar\sigma)}_{ba}(-\omega),
 \ee
 with the corresponding spectral functions of
 \be \label{Cwdef}
  C_{ab}^{(\sigma)}(\omega)
    \equiv \frac{1}{2}\int_{-\infty}^{\infty}dt
    e^{i\omega t}C^{(\sigma)}_{ab}(t).
 \ee
 The spectrum positivity that reads
 \be\label{positivity}
  C^{(\sigma)}_{aa}(\omega)\geq 0, \ \ \
  C^{(\sigma)}_{aa}(\omega)C^{(\sigma)}_{bb}
  (\omega) \geq |C^{(\sigma)}_{ab}(\omega)|^2,
 \ee
can be readily verified via the same method
used in the Appendix A of Ref.\ \onlinecite{Yan05187}.
   The symmetry and detailed-balance relations
 [\Eqs{Ctsym} or \Eqs{Cwsym}] are general
 for a grand ensemble, no matter it is of Fermion or Boson
 in nature.
  This nature enters in terms of FDT
 via the spectral density functions as follows.

   Consider $h_{\B}=\sum_k \epsilon_k c^{\dg}_kc_k$
 and $\hat f_a=\sum_k t_{ak}c_k$. In this case,
 the interaction spectral density functions are defined physically
 for $\omega\geq \min (\epsilon_k)\equiv 0$ as
 \be \label{Jwdef0}
   J_{ab}(\omega\geq 0) =
     \pi\sum_{k}  t_{ak}t^{\ast}_{bk} \delta(\omega-\epsilon_k)
  = J^{\ast}_{ba}(\omega).
 \ee
 Mathematically, one can extend their definition to
 the $\omega<0$ region, by setting
 $J_{ab}(\omega)\equiv \pm J_{ba}(-\omega)$
 for the Fermion/Boson ($+/-$) particles for the reasons that
 will become self-evident soon. As a result,
 \be\label{Jwdef1}
    J_{ab}(\omega) =
     \pi\sum_{k} [t_{ak}t^{\ast}_{bk} \delta(\omega-\epsilon_k)
     \pm t^{\ast}_{ak}t_{bk}\delta(\omega+\epsilon_k)].
 \ee
 On the other hand,
\bsube \label{ffcomm}
 \bea
   \bigl[\hat f_a(t),\hat f^{\dg}_b(0)\bigr]_{\pm}
 \!\!&=&\!\!
  \sum_{k} t_{ak}t^{\ast}_{bk} e^{-i\epsilon_kt},
 \label{ffcomm1} \\
   \bigl[\hat f^{\dg}_a(t), \hat f_b(0)\bigr]_{\pm}
 \!\!&=&\!\!
  \pm \sum_{k} t^{\ast}_{ak}t_{bk}e^{i\epsilon_kt},
 \label{ffcomm2}
\eea
\esube
 assuming $c$-numbers in this case.
 Consequently, \Eq{Jwdef1} can be recast as
 \bea \label{app_Jwdef2}
    J_{ab}(\omega)
  \!\!&=&\!\! \frac{1}{2}
    \int_{-\infty}^{\infty}\!\!dt e^{i\omega t}
     \bigl\la\bigl[\hat f_a(t),\hat f^{\dg}_b(0)\bigr]_{\pm}\bigr\ra_{\B}
\nl \!\!&=&\!\! \frac{1}{2}
    \int_{-\infty}^{\infty}\!\!dt e^{i\omega t}
     \bigl\la\bigl[\hat f^{\dg}_a(t),\hat f_b(0)\bigr]_{\pm}\bigr\ra_{\B},
 \eea
 or [cf.\ \Eqs{ffcorr0} and (\ref{Cwdef})]
 \be \label{Jwdef}
   J_{ab}(\omega) = C^{(\sigma)}_{ab}(\omega)
    \pm C^{(\bar\sigma)}_{ba}(-\omega).
 \ee
 Note the symmetry relations here,
\be \label{Jwsym}
   J_{ba}(\omega)=\pm J_{ab}(-\omega)  =J^{\ast}_{ab}(\omega).
\ee
 From the second identity of \Eq{Cwsym}, we have also
 \be\label{FDTw}
   C_{ab}^{(\sigma)}(\omega)
   =\frac{J_{ab}(\omega)}{1\pm e^{-\beta(\omega+\sigma\mu)}},
 \ee
 or equivalently
 \be \label{FDTt}
 C^{(\sigma)}_{ab}(t) = \frac{1}{\pi} \int_{-\infty}^{\infty}\!d\omega
     \frac{e^{-i \omega t}J_{ab}(\omega)}
     {1\pm e^{-\beta(\omega+\sigma\mu)}}  .
 \ee
  This is the FDT for the Fermion ($+$) or Boson ($-$) grand
 canonical bath ensembles.

   In concluding this section, let us make some comments
 on \Eq{Jwdef}, which  can in fact be considered as the
 working definition of the spectral density functions in
 general. For the linear bath interaction model,
 \Eq{Jwdef} is equivalent to the conventional
 expression, \Eq{Jwdef0} or \Eq{Jwdef1}, in which
 $J_{ab}(\omega)$ is independent of temperature and
 chemical potential.
 As the working definition, \Eq{Jwdef} may also
 be useful even for the cases where
 the bath interaction depends nonlinearly on
 the annihilation (creation)
 operators of the Fermion/Boson bath ensembles.
 In such cases, \Eq{Jwdef} that does not depend on
 the $\sigma$-index may involve with
  a certain mean-field approximation and the
 resulting $J_{ab}(\omega)$
 remains explicitly independent of temperature
 and chemical potential.
 The implicit dependence of $J_{ab}(\omega)$ on
 temperature and/or chemical potential
 may however arise from the mean-field
 values of $t_{ak}$ or $t_{bk}$, as
 in \Eq{Jwdef0} or (\ref{Jwdef1}).

 Consider now the value of $J_{ab}(\omega=0)$.
 As inferred from the second identity of \Eq{Jwsym},
 $J_{ab}(0) =\pm J^{\ast}_{ab}(0)$,
 implying that $J_{ab}(0)$ is pure real/imaginary for
 a Fermion/Boson bath interaction.
  For the spectral density function, assuming it is
  analytical at $\omega=0$,
 the Boson bath interaction will also assume
 \[
 J_{ab}(0)=
  \bigl[
   (1-e^{-\beta\omega}) C^{(\sigma)}_{ab}(\omega)
  \bigr]_{\omega\rightarrow 0;\mu=0} = 0;   \ \ ({\rm Boson})
 \]
 as inferred from \Eq{FDTw} and from the fact
 that $J_{ab}(\omega)$ does not depend
 on $\mu$ explicitly.

  We are interested in nonperturbative QDT,
 in the presence of non-Markovian multi-mode interaction
 with grand canonical Fermion/Boson bath.
 Both the path integral and the differential EOM
 formalisms of QDT will be considered.
 The former will be presented in the coming
 section, while the latter will be constructed
 in \Sec{thgeneral}, on the basis of calculus
 (time-derivative) on the exact path integral expressions.
  In order to have the EOM hierarchically coupled and closed
 for a general non-Markovian case,
 the bath correlation functions $C^{(\sigma)}_{ab}(t)$
 should be parameterized in a proper form.
 The aforementioned properties of $J_{ab}(\omega=0)$
 will be used there, together with \Eqs{Jwsym} and \Eq{FDTt},
 to the FDT-preserved parametrization of $C^{(\sigma)}_{ab}(t)$
 in \Sec{thparameter}
 that participates in the construction of the hierarchical EOM
 for general non-Markovian dissipation dynamics.

\section{Path integral formalism and dissipation functionals}
\label{thpath}

 In this section, we
  construct the
 path integral formalism of quantum dissipation.
 For the sake of clarity, we present only the
 final results here.
 The derivation of the path integral
 formalism will be given in \App{thapp_path}.
 It is similar to our previous work
 on the Bosonic canonical bath interaction cases,
 where the system and bath operators,
 $\{W_a\}$ and $\{f_a\}$, were also assumed to be
 Hermitian.\cite{Xu05041103,Xu07PRE}
 Briefly speaking, it employs
 the Wick's theorem for the thermodynamic
 Gaussian average, followed by using the
 symmetry and detailed--balance relations
 of \Eq{Ctsym}.  It involves also the ansatz of
 the initially factorizable total density operator,
 $\rho_{\rm T}(t_0)=\rho(t_0)\rho^{\rm eq}_{\B}$,
 with $\rho^{\rm eq}_{\B}$ being given by \Eq{rhoeq}.
 Note that when the initial time $t_0\rightarrow-\infty$
 this ansatz imposes no approximation.\cite{Wei99,Yan05187}

  Unlike its EOM counterpart that can be expressed in
 the operator level, the path integral formalism goes
 with a representation. Let $\{|\alpha\ra\}$ be a basis
 set in the system subspace, and
 $\bfalp \equiv (\alpha,\alpha')$ for abbreviation,
 so that $\rho(\bfalp,t)\equiv \rho(\alpha,\alpha',t)$.
 Denote ${\cal U}(t,t_0)$ as the reduced
 Liouville--space propagator,
 $\rho(t) \equiv {\cal U}(t,t_0)\rho(t_0)$,
 which in the $\alpha$-representation reads
 \be \label{rhotPI_def}
 \rho({\bm \alpha},t) = \!\int\!d\bfalp_0\,
  {\cal U}(\bfalp,t;\bfalp_0,t_0) \rho(\bfalp_0,t_0),
 \ee
 with the path integral expression of \cite{Fey63118}
 \be \label{calGPI}
   {\cal U}(\bm\alpha,t;\bm\alpha_0,t_0)
 = \int_{\bm\alpha_0[t_0]}^{\bm\alpha[t]}   \!\!  {\cal D}{\bm\alpha} \,
     e^{iS[\alpha]} {\cal F}[\bm\alpha] e^{-iS[\alpha']}.
 \ee
 $S[\alpha]$ is the classical action functional
 of the reduced system, evaluated along a path $\alpha(\tau)$, with the
 constraints that two ending points $\alpha(t_0)=\alpha_0$ and
 $\alpha(t)=\alpha$ are fixed.
 The key quantity in the path integral formalism
  is the Feynman--Vernon influence
 functional ${\cal F}[\bfalp]$ that will be presented soon.

   In connection to the later development of EOM formalism,
we denote hereafter ${\bm a}=(aa')$ for a pair of dissipation modes
and introduce
 \be  \label{ticalW_PI}
   \ti{\cal W}^{(\sigma)}_{\bm a}(t;\{\bm\alpha\})
  \equiv
  \ti W^{(\sigma)}_{aa'}(t;\{\alpha\}) -
  \ti W^{\prime(\sigma)}_{aa'}(t;\{\alpha'\}),
 \ee
 where \big[noting that $\ti W^{(\sigma)}_{\bm a}\equiv \ti W^{(\sigma)}_{aa'}$
 and $C^{(\sigma)}_{\bm a}\equiv C^{(\sigma)}_{aa'}$\big]
 \bsube \label{tiW_PI}
 \bea \label{tiWa_PI}
 \ti W^{(\sigma)}_{\bm a}(t;\{\alpha\})
 &\!\!\equiv&\!\!\!\int_{t_0}^{t}\!\!d\tau
    C^{(\sigma)}_{\bm a}(t-\tau) W^{\sigma}_{a'}[\alpha(\tau)],
 \\
 \label{tiWpa_PI}
   \ti W^{\prime(\sigma)}_{\bm a}(t;\{\alpha'\})
 &\!\!\equiv&\!\!\!\int_{t_0}^{t}\!\!d\tau
    [C^{(\bar\sigma)}_{\bm a}(t-\tau)]^{\ast}
    W^{\sigma}_{a'}[\alpha'(\tau)].
 \eea
 \esube
 Denote also
 \be \label{calW_PI}
    {\cal W}^{\sigma}_a[\bfalp] \equiv
     W^{\sigma}_a[\alpha(t)]-W^{\sigma}_a[\alpha'(t)]
  \equiv W^{\sigma}_a - W^{\prime\sigma}_a.
 \ee
 It is in fact the $W^{(\sigma)}_a$--commutator
 in the path-integral representation, as it depends
 only at the fixed ending points of the path.
 The time-dependence here can thus be removed.

 The final expressions for the influence functional are
 [cf.\ \Eq{app_calR_PI}]
 \bsube\label{calFfinal}
 \be\label{FinR}
    {\cal F}[\bm\alpha]\equiv
   \exp \left\{-\int^t_{t_0}\!d\tau{\cal R}[\tau;\{\bm\alpha\}]\right\},
  \ee
  with
  \be\label{calR_PI}
   {\cal R}[t;\{\bm\alpha\}]\equiv\sum_{{\bm a},\sigma}
    {\cal W}^{\bar\sigma}_a[\bfalp]
    \ti{\cal W}^{(\sigma)}_{\bm a}(t;\{\bm\alpha\}).
 \ee
 \esube
 Here, ${\cal F}$ is expressed in terms of its exponent
 time-integrand ${\cal R}$.
 It is in fact the {\it time-local dissipation superoperator}
 in the path integral representation, as
 the time derivative of \Eq{FinR},
 \be\label{dotFinR}
   \partial_t {\cal F} = -{\cal RF},
 \ee
 leads to
 $\partial_t {\cal U} = -i{\cal L}\,{\cal U} -{\cal R}(t){\cal U}$
 [cf.\ \Eq{calGPI}],
 or equivalently
 $\dot{\rho} = -i{\cal L}\,\rho -{\cal R}(t)\rho$
 [cf.\ \Eq{rhotPI_def}].
 Therefore, ${\cal R}$ of \Eq{calR_PI} can be termed
 as the {\it dissipation functional}.

 It is noticed that the chemical potential
 $\mu$ does not appear explicitly in the
 dissipation functional ${\cal R}$.
 The above exact path integral formalism,
 \Eqs{rhotPI_def}--(\ref{calFfinal}),
 is formally the same for both the canonical
 and the grand canonical bath interactions,
 no matter the bath is of Fermion or Boson,
 as long as the Gaussian bath statistics is applicable.
 Apparently, in the case of $W_{a}=W^{\dg}_a\equiv Q_a$
 and $\hat f_{a}=\hat f^{\dg}_a=\hat F_a$, the
 sign-index $\sigma$ ($``+"$ and $``-"$)
 can be omitted, and \Eqs{ticalW_PI}--(\ref{calFfinal})
 recovers formally the Eqs.\,(10)--(13) of Ref.\ \onlinecite{Xu07PRE},
 where the Bosonic canonical bath ensembles were also assumed.
 The different nature of the quantum bath ensemble
 is embedded implicitly
 in the correlation functions $C^{(\sigma)}_{\bm a}(t)$,
 via the FDT and symmetry relations
 as described in \Sec{thmodelB}; see \Eqs{Jwsym} and (\ref{FDTt}).

\section{Non-Markovian bath via parameterization}
\label{thparameter}
   To construct a hierarchically coupled set of
 EOM 
 for a general non-Markovian dissipation,
 $C^{(\sigma)}_{\bm a}(t)$ should be expressed in a proper form,
 such as an exponential series
 expansion,\cite{Wel06044712,Xu05041103,Xu07PRE,Mei993365}
 which shall also satisfy the FDT [\Eq{FDTt}].
  To that end, let us consider the
 following parametrization form on the
 bath spectral density functions ($J_{\bm a}\equiv J_{aa'}$),
 \bsube\label{Jw_para}
 \bea\label{Jw_paraA}
  J_{\bm a}(\omega)
 \!\!&=&\!\!
  J^{\bm a}_{\D}(\omega) +
 \sum^K_{k=1}\Bigg[
 \frac{\zeta^{\bm a}_k\gamma^{\bm a}_k+i\bar\zeta^{\bm a}_k\omega}
    {(\omega-\omega^{\bm a}_k)^2+(\gamma^{\bm a}_k)^2}
 \nl \!\!&\ &\!\! \qquad
 \pm\,
  \frac{\zeta^{\bm a}_k\gamma^{\bm a}_k+i\bar\zeta^{\bm a}_k\omega}
    {(\omega+\omega^{\bm a}_k)^2+(\gamma^{\bm a}_k)^2}
  \Bigg],
 \eea
 with the Drude term
 \be \label{Jw_drude}
  J^{\bm a}_{\D}(\omega)=\frac{\zeta^{\bm a}_{\D}\gamma^{\bm a}_{\D}
  +i\bar\zeta^{\bm a}_{\D}\omega}{\omega^2+(\gamma^{\bm a}_{\D})^2}
 \quad {\rm or}\quad
    \frac{\zeta^{\bm a}_{\D}\omega}{\omega^2+(\gamma^{\bm a}_{\D})^2} .
 \ee
 \esube
 for the Fermion ($+$) or Boson ($-$)
 bath ensembles, respectively.

 All involving parameters are real;
 $\omega^{\bm a}_k$ and $\gamma^{\bm a}_k$
 (including $\gamma^{\bm a}_{\D}$)
 are positive as well. They satisfy [cf.\ \Eq{Jwsym}]
 $(\omega^{aa'}_{k},\gamma^{aa'}_{k},
 \zeta^{aa'}_{k},\bar\zeta^{aa'}_{k})
  =(\omega^{a'\!a}_{k},\gamma^{a'\!a}_{k},
 \zeta^{a'\!a}_{k},-\bar\zeta^{a'\!a}_{k})$.
 The corresponding bath correlation
 functions can then be obtained via the FDT of \Eq{FDTt},
 using the contour integration method.
  The final results read ($M\rightarrow\infty$)
 \be \label{corr_para}
   C^{(\sigma)}_{\bm a}(t)
= 
   \eta^{(\bm a)}_{\D}e^{-\gamma^{\bm a}_{\D}t} 
 +\sum_{j=1}^{2K}\eta^{(\bm a)}_{j}\phi^{\bm a}_j(t)
 +\sum^{M}_{m=1}\check C^{(\bm a)}_{m}(t).
 \ee
 Hereafter, $\eta^{(\bm a)}_j\equiv \eta^{(\sigma)}_{\bm a,j}$
 and $\eta^{(\bar{\bm a})}_j\equiv \eta^{(\bar\sigma)}_{\bm a,j}$
 and similarly for other parameters
 that depend both $(\sigma)$ and ${\bm a}$.

 The first two terms in the rhs of \Eq{corr_para}, where
 \bsube\label{phik}
 \bea
  \phi^{\bm a}_{2k-1}(t)
 \!\!&\equiv&\!\!
  \sin(\omega^{\bm a}_k t) \exp(-\gamma^{\bm a}_kt),
\\
  \phi^{\bm a}_{2k}(t)
 \!\!&\equiv&\!\!
  \cos(\omega^{\bm a}_k t)\exp(-\gamma^{\bm a}_kt),
 \eea
 \esube
  arise from the poles of $J_{\bm a}(z)$
 in \Eq{Jw_para}.
 The involving $\eta$--coefficients are
\[ 
   \eta^{(\bm a)}_{\D}
  = \frac{\zeta^{\bm a}_{\D}+\bar\zeta^{\bm a}_{\D}}
   {e^{i\beta(\gamma^{\bm a}_{\D}+\sigma i\mu)}+1}
 \ \ \text{or}\ \
   \frac{i\zeta^{\bm a}_{\D}}
 {e^{i\beta(\gamma^{\bm a}_{\D}+\sigma i\mu)}-1},
\] 
 for the Fermion ($+$) or Boson ($-$), respectively;
$ \eta^{(\bm a)}_{2k-1}
 = -i\lambda^{\bm a}_k \bar \nu^{(\bm a)}_{k}
  \pm i \bar \lambda^{\bm a}_k \nu^{(\bm a)}_{k}$
and
$\eta^{(\bm a)}_{2k}
 = \lambda^{\bm a}_k \bar\nu^{(\bm a)}_{k}
  \pm \bar \lambda^{\bm a}_k \nu^{(\bm a)}_{k}$.
Here,
 $\lambda^{\bm a}_k \equiv[(\zeta^{\bm a}_k
  +\bar\zeta^{\bm a}_k)\gamma^{\bm a}_k
 +\bar\zeta^{\bm a}_k\omega^{\bm a}_k]/(2\gamma^{\bm a}_k)
 \equiv (\bar\lambda^{\bm a}_k)^{\ast}$,
 while $\nu^{(\bm a)}_k\equiv
  \{1\pm \exp[\beta(\omega^{\bm a}_k+i\gamma^{\bm a}_k-\sigma\mu)]\}^{-1}$,
 and $\bar\nu^{(\bm a)}_k$ is similar as $\nu^{(\bm a)}_k$
 but with $-\omega^{\bm a}_k$ instead.

  The last term of \Eq{corr_para}
 arises from the poles of Matsubara frequencies,
 \be \label{checkgam}
   \check\gamma_m \equiv (2m-1)\pi/\beta
   \quad {\rm or}\quad 2m\pi/\beta
 \ee
 for the Fermion or Boson bath ensemble, respectively.
 \be \label{checkCm}
  \check C^{(\bm a)}_{m}(t)\equiv
   \check\eta^{(\bm a)}_{m}
   \exp\left[-(\check{\gamma}_m-\sigma i \mu)t\right]
 =\mp [\check C^{(\bar{\bm a})}_{m}(t)]^{\ast},
 \ee
 with
 \be\label{checketasym}
  \check\eta^{(\bm a)}_m \equiv
  -i(2/\beta)J_{\bm a}(-i\check\gamma_m-\sigma\mu)=
  \mp [\check\eta^{(\bar{\bm a})}_m]^{\ast}.
 \ee
 The last identity in each of the above two equations
 is originated from
 $J_{\bm a}(-z)=\pm [J_{\bm a}(z)]^{\ast}$,
 the analytical continuation of
 $J_{\bm a}(-\omega)=\pm J^{\ast}_{\bm a}(\omega)$ [cf.\ \Eqs{Jwsym}].
  It will show that this property leads to some simplification
 in treating the Matsubara-frequency contributions.
 In the canonical ensemble,
 the Matsubara coefficients $\check\eta^{(\bm a)}_m$
 of \Eq{checketasym}, which do not depend on $\sigma$ and
 chemical potential $\mu$, are real and pure imaginary for Boson
 and Fermion bath ensembles, respectively.

  To conclude this section, let us make some comments
 on the parameterized $C^{(\sigma)}_{\bm a}(t)$ of \Eq{corr_para},
 for which the hierarchical EOM formalism
 can be constructed via the IGF-COPI method\cite{Xu05041103,Xu07PRE}
 without approximations  (cf.\ \Sec{thgeneral}).
 In principle, \Eq{corr_para} can be exact for a general
 non-Markovian dissipation, if the involving
 $K$ and $M$ are sufficiently large.
 The resulting EOM formalism is also exact;
 but its size grows in a power law as $K$ or $M$ increases.
 The exact evaluation of complex dissipation
 would rapidly become extremely tedious.
 In practise, the $K$ and $M$ in \Eq{corr_para}
 have to be finite in the construction of the EOM
 formalism.
 To complete the theory, the residue correction due to
 the difference between the true and
 the parameterized correlation functions,
 $\delta C^{(\sigma)}_{\bm a}\equiv
  C^{(\sigma)}_{\bm a,\rm exa}-C^{(\sigma)}_{\bm a}$,
 should also be incorporated into the final EOM formalism.

 \section{Approach to the exact equations
    of motion formalism: Principles}
 \label{thspecial}

 \subsection{Hierarchy construction}
 \label{thspecialA}

   For the sake of clarity,
 we exploit the broadband dissipation case
 to illustrate the IGF-COPI approach
 to the hierarchical EOM, together
 with the residue correction and hierarchy truncation
 that will be demonstrated in the next subsection.
 In the broadband limit,
 $\eta^{(\bm a)}_{\D}\exp(-\gamma^{\bm a}_{\D}t)
 \rightarrow \eta^{(\bm a)}_{\D}\gamma^{\bm a}_{\D}\delta(t)$
 and the $\eta^{(\bm a)}_{j}$--term
 in \Eq{corr_para} can be completely neglected.
 We will show in \Sec{thresidue}
 that the $\delta(t)$-like component
 can be easily taken into account in terms
 of the residue correction.

  Considered here explicitly for the illustration
 of hierarchy construction
 are only the Matsubara terms,
 \be\label{corrm}
   C^{(\sigma)}_{\bm a}(t) = \sum_{m=1}^M \check C^{(\bm a)}_m(t)
   = \sum_{m=1}^M \check\eta^{(\bm a)}_m
  e^{-(\check\gamma_m-\sigma i\mu)t}.
 \ee
 The dissipative functional ${\cal R}$  [\Eq{calR_PI}],
 as it appears linearly in the bath correlation functions,
 can be expanded like \Eq{corrm} as
 \be\label{calR_mar0}
   {\cal R}=\sum_{\bm a,\sigma,m}{\cal W}^{\bar\sigma}_a[\bfalp]
     \check{\cal W}^{(\bm a)}_m(t;\{\bfalp\}) .
 \ee
 Here, $\check{\cal W}^{(\bm a)}_m$ are
 defined by \Eqs{ticalW_PI} and (\ref{tiW_PI})
 but with the bath correlation functions being
 replaced by $\check C^{(\bm a)}_m$.
 We shall hereafter omit the explicit path-integral variables
 dependence whenever it does not cause confusion.

   Note that the Matsubara terms
 possess two special properties:
 $\check C^{(\bm a)}_m(t)
 = \pm [\check C^{(\bar{\bm a})}_m(t)]^{\ast}$ [cf.\ \Eq{checkCm}]
 and the fact that their time constants
 $(\hat \gamma_{m}-\sigma i\mu)$ are dissipation-mode independent.
 As a result, \Eq{calR_mar0} can have
 the summation over $a'$ performed, resulting in
 \be\label{calR_mar}
   {\cal R} = i\sum_{a,\sigma,m}
   {\cal W}^{\bar\sigma}_a \check Z^{(a)}_{m}.
 \ee
 The involving \Eqs{ticalW_PI} and (\ref{tiW_PI}) are rearranged
 for the Matsubara terms as
 \be\label{checkZm}
   \check Z^{(a)}_m
 \equiv -i\sum_{a'} \check{\cal W}^{(\bm a)}_{m}
  = -i\left(\check W^{(a)}_{m} \pm \check W^{\prime(a)}_{m}\right),
 \ee
 with
 \bsube \label{checkWm}
 \bea
  {\check W}^{(a)}_{m}
 \!\!&\equiv&\!\!
  \sum_{a'}\int_{t_0}^{t}\!d\tau\,
  \check C^{(\bm a)}_{m}(t-\tau)W^{\sigma}_{a'}[\alpha(\tau)],
 \\
  {\check W}^{\prime(a)}_{m}
 \!\!&\equiv&\!\!
  \sum_{a'} \int_{t_0}^{t}\!d\tau\,
    \check C^{(\bm a)}_{m}(t-\tau)W^{\sigma}_{a'}[\alpha'(\tau)].
 \eea
 \esube
 We have
 \be\label{dotcheckZm}
  \partial_t\check{Z}^{(a)}_m
  = \check{\cal C}^{(a)}_m -(\hat\gamma_m-\sigma i\mu)
   \check{Z}^{(a)}_m,
 \ee
 where
 \be \label{checkcalCm}
  \check{\cal C}^{(a)}_m\equiv
 -i\sum_{a'} \check\eta^{(\bm a)}_m
   \left(W^{\sigma}_{a'} \pm W^{\prime\,\sigma}_{a'}\right).
 \ee

  We are now in the position to construct
 the EOM via the calculus (time-derivative) on the
 path integral formalism.\cite{Xu05041103,Xu07PRE}
 The time derivative on the action functional parts contributes to
 the coherent dynamics of $-i{\cal L}U$, and thus can be included
 into the final EOM formalism. However, the time derivative of
  the influence functional, \Eq{dotFinR}, will
 generate other {\it auxiliary influence functionals}
 (AIFs). This process goes on  progressively and hierarchically
 as the time derivatives on AIFs is continued.
 It is also noticed that time derivative of individual
 $\check Z^{(a)}_m$ is closed by itself [\Eq{dotcheckZm}].
 We will see soon that this property,
 together with their exclusive relation to
 the dissipation functional as \Eq{calR_mar},
 make $\{\check Z^{(a)}_m\}$
 the complete set of the 
 IGFs for the desired hierarchy construction.

  Let us start with the time derivative on the
 primary influence functional, \Eq{dotFinR}. In the present
 demonstrative case, it reads
 \be\label{dotFmar}
   \partial_t{\cal F}=-{\cal RF} = -i\sum_{a,\sigma,m}
    {\cal W}^{\bar\sigma}_a{\cal F}^{(a)}_m.
 \ee
 Generated here are the first-tier AIFs, defined as
 \be\label{F1mar}
   {\cal F}^{(a)}_m \equiv \check{Z}^{(a)}_m{\cal F}.
 \ee
 Compared with the primary ${\cal F}$ [\Eq{calFfinal}] whose leading
 term is 1, the first-tier AIFs $\{{\cal F}^{(a)}_m\}$
 are of the second-order in the system--bath coupling
 as their leading contributions.

  To proceed, the time derivatives on
 those first-tier AIFs are also needed.
 We obtain [cf.\ \Eqs{dotcheckZm}--(\ref{F1mar})],
 \be \label{dotF1mar}
   \partial_t{\cal F}^{(a)}_m
 =
   \left[\check{\cal C}^{(a)}_m{\cal F}
    -(\hat\gamma_m-\sigma i\mu){\cal F}^{(a)}_m\right]
   + \check{Z}^{(a)}_m (\partial_t{\cal F}).
 \ee
 Here,
 \be\label{ZmdotFmar}
   \check{Z}^{(a)}_m (\partial_t{\cal F})
  = -i\sum_{a',\sigma',m'}
    {\cal W}^{\bar{\sigma}'}_{a'}{\cal F}^{(aa')}_{mm'},
 \ee
 with (denoting $\check{Z}^{(a')}_{m'}
       \equiv \check{Z}^{(\sigma')}_{a',m'}$)
 \be \label{F2mar}
  {\cal F}^{(aa')}_{mm'} \equiv \check{Z}^{(a)}_m
   {\cal F}^{(a')}_{m'}
 = \big(\check{Z}^{(a)}_m\check{Z}^{(a')}_{m'}\big){\cal F},
 \ee
 being the next tier (second-tier) AIFs.
 Apparently, the leading contributions in
 ${\cal F}^{(aa')}_{mm'}$ are of the fourth-order
 in the system--bath coupling.

  Continue on applying \Eqs{dotcheckZm}--(\ref{dotFmar})
  for \Eq{F2mar}
 \be\label{dotF2}
  \partial_t{\cal F}^{(aa')}_{mm'} =
   \big[\partial_t\big(\check{Z}^{(a)}_m\check{Z}^{(a')}_{m'}
   \big)\big]{\cal F}
  + \big(\check{Z}^{(a)}_m\check{Z}^{(a')}_{m'}\big)
    \big(\partial_t{\cal F}\big).
 \ee
 The first term in the rhs above will lead to
 the dependence on the tier-minus-one AIFs,
 arising from the $\check{\cal C}^{(a)}_m$--term of \Eq{dotcheckZm},
 while the $(\partial_t{\cal F})$--term above leads to
 the tier-plus-one AIFs.

  Apparently, those distinct $\{\check Z^{(a)}_{m}\}$
 are IGFs, since all involving AIFs can be generically expressed as
 \be\label{AIFmar0}
   {\cal F}^{(aa'\cdots a'')}_{mm'\cdots m''}
  = \big(\check{Z}^{(a)}_m\check{Z}^{(a')}_{m'}
   \cdots\check{Z}^{(a'')}_{m''}\big){\cal F}.
 \ee
 In order to eliminate possible duplications due to the indexes
 permutation/repetition that result in the same AIF,
 we arrange all distinct $\{\check{Z}^{(a)}_m\}$
 in a specified sequence and rearrange \Eq{AIFmar0} accordingly as
 \be\label{AIFmar}
    {\cal F}_{\ind}
  \equiv
  \Big\{\prod_{a,\sigma,m}
   \big(\check Z^{(a)}_{m}\big)^{\check n^{(a)}_m}\Big\}
  {\cal F}.
 \ee
 Here ${\sf n}=\{\check n^{(a)}_m\}$ is the
 index-set of nonnegative integers
 for the number of occurrences on the distinct
 $\check Z^{(a)}_{m}$ individuals in the specified sequence.
 Note that the total number of nonnegative integers in
 the index-set ${\sf n}$ is $2Mq$, where
 $M$ is the number of Matsubara terms considered,
 $q$ the number of dissipative mode $a$, while
 the factor 2 arises from $\sigma=+,-$.

   To specify the hierarchical dependence
 of $\partial_t{\cal F}_{\sf n}$,
  we denote also the index-set
 $\check{\sf n}^{\pm}_{(a),m}
  \equiv \{\cdots, {\check n}^{(a)}_m \pm 1,\cdots\}$
 that deviates from ${\sf n}$ only by
 changing the specified $\check n^{(a)}_m$ to $\check n^{(a)}_m\pm 1$.
  The time derivative of ${\cal F}_{\sf n}$ [\Eq{AIFmar}]
 can be carried out, again by
 using \Eqs{dotcheckZm} and (\ref{dotFmar}), resulting in
 \bea \label{dotFn_m}
   \partial_t{\cal F}_{\sf n}
 \!\!&=&\!\!
  -(\check\gamma_{\sf n}+i\check n\mu){\cal F}_{\sf n}
     +\sum_{a,\sigma,m}
  \check n^{(a)}_m\check{\cal C}^{(a)}_{m}
  {\cal F}_{\check{\ind}_{(a)}^{-}}
\nl\!\!&\ &\!\!
    -i\sum_{a,\sigma,m}{\cal W}^{\bar\sigma}_a
    {\cal F}_{\check{\ind}^+_{(a),m}} .
 \eea
 Here,
 \bsube\label{checkgam_n_mar}
 \bea
   \check\gamma_{\sf n}
 \!\!&\equiv&\!\!
   \sum_{a,\sigma,m}
   \check n^{(a)}_m \check\gamma_m
  =
   \sum_{a,m} [\check n^{(-)}_{a,m}+\check n^{(+)}_{a,m}]\check\gamma_m,
 \\ 
 \label{checkn_mar}
  \check n
  \!\!&\equiv&\!\!
    -\sum_{a,\sigma,m} \sigma\check n^{(a)}_m
  = \sum_{a,m} [\check n^{(-)}_{a,m} - \check n^{(+)}_{a,m}].
 \eea
 \esube

  The auxiliary reduced density operators can now be defined as
(including also $\rho\equiv\rho_{\sf 0}$)
 \be \label{rhon_def}
  \rho_{\sf n}(t)\equiv {\cal U}_{\sf n}(t,t_0)\rho(t_0),
 \ee
 with the auxiliary propagators of [cf.\ \Eq{calGPI}]
 \bea \label{calUn_def}
   {\cal U}_{\sf n}(\bfalp,t;\bfalp_0,t_0)
 \equiv \int_{\bfalp_0}^{\bfalp}   \!\!  {\cal D}{\bfalp} \,
     e^{iS[\alpha]} {\cal F}_{\sf n}[\bfalp] e^{-iS[\alpha']} .
 \eea
 The hierarchically coupled \Eqs{dotFn_m} can now be recast as
 \bea \label{dotrhon_mar}
   \dot{\rho}_{\sf n}
 \!\!&=&\!\!
  -\big[i({\cal L}+\check n\mu) +\check\gamma_{\sf n}\big]\rho_{\sf n}
     +\sum_{a,\sigma,m}
  \check n^{(a)}_m\check{\cal C}^{(a)}_{m}
  {\rho}_{\check{\ind}_{(a)}^{-}}
\nl\!\!&\ &\!\!
    -i\sum_{a,\sigma,m}{\cal W}^{\bar\sigma}_a
    {\rho}_{\check{\ind}^+_{(a),m}} .
 \eea
Here, ${\cal W}^{\bar\sigma}_a$ and $\check{\cal C}^{(a)}_{m}$,
 which were given in the path integral representation
 as \Eqs{calW_PI} and (\ref{checkcalCm}),
 respectively, are now the reduced Liouville-space operators; i.e.,
\bea \label{calWcheckC}
 {\cal W}^{\sigma}_a\hat O =[W^{\sigma}_a,\hat O],
 \ \ \
 \check{\cal C}^{(a)}_{m}\hat O
 = -i\sum_{a'}\check\eta^{(\bm a)}_m[W^{\sigma}_{a'},\hat O]_{\pm}.
\eea

\subsection{Residue correction and the hierarchy truncation}
\label{thresidue}
    We have illustrated the IGF-COPI approach to
 the hierarchical EOM formalism. The same method will be
 used in \Sec{thgeneral} to establish the EOM formalism
 for a general non-Markovian dissipation characterized
 by \Eq{corr_para} without approximation.
 The final hierarchical EOM [cf.\ \Eq{dotrhon}]
 can be expressed as
 \be \label{dotrhon0}
 \dot\rho_{\ind}=-[i({\cal L}+\check n \mu)
   +{\mathit\gamma_{\ind}}+\delta{\cal R}(t)]\rho_{\ind}
  +\rhonswap +\rhondown + \rhonup.
 \ee
 The hierarchy-down $\rhondown$ and hierarchy-up $\rhonup$
 are similar as the last two terms in \Eq{dotrhon_mar},
 but with the Drude and the $\phi^{\bm a}_j$--components of the bath
 correlation functions of \Eq{corr_para} included.
 The hierarchical-swap $\rhonswap$
 arises completely from the $\phi^{\bm a}_j$--components,
 thus have no correspondence in \Eq{dotrhon_mar}.

  Involved in \Eq{dotrhon0} is also the
 residue correction $\delta{\cal R}(t)$ to the dissipation.
 It accounts for the practical difference
 between the exact $C^{(\sigma)}_{{\bm a},{\rm exa}}(t)$
 and the parameterized $C^{(\sigma)}_{\bm a}(t)$ by
 \Eq{corr_para} where $K$ and $M$ are finite.
 In the following we shall present the principle
 of residue correction\cite{Xu07PRE} to construct the residue
 dissipation $\delta{\cal R}$, followed by
 an efficient scheme of hierarchy truncation.

  Let us first verify that
  the residue dissipation correction
  $\delta{\cal R}={\cal R}_{\rm exa}-{\cal R}$,
 due to $\delta C^{(\sigma)}_{\bm a}
  \equiv C^{(\sigma)}_{{\bm a},{\rm exa}}
   - C^{(\sigma)}_{\bm a}$,
 does enter at the global
 level, participating in every tier of the hierarchy
 as \Eq{dotrhon0}.
 As inferred from \Eq{FinR},
 the exact influence functional of primary interest reads in
 path integral representation as
 ${\cal F}^{\rm exa} = {\cal F}\,{\cal F}_{\rm resi}$,
 with $\partial_t{\cal F}=-{\cal RF}$ and
 $\partial_t {\cal F}_{\text{resi}} =
  -\delta{\cal R}\, {\cal F}_{\text{resi}}$.
  The auxiliary influence functionals defined, for example in
 \Eq{AIFmar}, for the construction of the hierarchical EOM
 should now be replaced by
 ${\cal F}^{\text{exa}}_{\ind}
  \equiv {\cal F}_{\ind}{\cal F}_{\text{resi}}$.
 Its time derivative is then
 $\partial_t {\cal F}^{\text{exa}}_{\ind}
 = (\partial_t{\cal F}_{\ind}){\cal F}_{\text{resi}}
   -{\delta\cal R} {\cal F}^{\text{exa}}_{\ind}$.
 Together with the established $\partial_t{\cal F}_{\ind}$
 such as \Eq{dotFn_m} before residue correction,
 the above expression verifies the fact that
 $\delta{\cal R}$ does enter into every tier of the
 final hierarchical EOM, as in \Eq{dotrhon0}.

  We are now in the position to present some practically
 useful expressions of $\delta{\cal R}(t)$.
 Note that its exact expression in path integral
 representation is [cf.\ \Eq{calR_PI}
 with \Eqs{ticalW_PI} and (\ref{tiW_PI})]
 \be \label{delta_calR}
  \delta{\cal R} =
  \sum_{{\bm a},\sigma}{\cal W}^{\bar\sigma}_a[\bfalp]\;
  \delta\ti{\cal W}^{(\sigma)}_{\bm a}(t;\{\bfalp\}),
 \ee
 where
 \be \label{del_ticalW}
   \delta{\ti{\cal W}}^{(\sigma)}_{\bm a}
   = \delta\ti W^{(\sigma)}_{\bm a}(t;\{\alpha\})
   - \delta\ti W^{\prime(\sigma)}_{\bm a}(t;\{\alpha'\}),
 \ee
 with
 \bsube\label{delta_tilWa}
 \bea
  \delta{\ti W}^{(\sigma)}_{\bm a}
 \!\!&=&\!\!
  \int_{t_0}^{t}\!d\tau\, \delta C^{(\sigma)}_{\bm a}(t-\tau)
   W^{\sigma}_{a'}[\alpha(\tau)],
 \\
  \delta{\ti W}^{\prime(\sigma)}_{\bm a}
 \!\!&=&\!\!
  \int_{t_0}^{t}\!d\tau\,
  \big[\delta C^{(\bar\sigma)}_{\bm a}(t-\tau)\big]^{\ast}
   \, W^{\sigma}_{a'}[\alpha'(\tau)].
 \eea
 \esube

  In the Markovian-residue limit, \Eqs{delta_tilWa} reduce to
  \be   \label{delta_tilWa_mar}
   \delta{\ti W}^{(\sigma)}_{\bm a}
 \approx
   \delta{\bar C}^{(\sigma)}_{\bm a}(t) W^{\sigma}_{a'}[\alpha(t)],
 \ee
 and similar for $\delta{\ti W}^{\prime(\sigma)}_{\bm a}$; with
 \be
   \delta{\bar C}^{(\sigma)}_{\bm a}(t)
   =\int_{t_0}^{t}\!d\tau\,\delta C^{(\sigma)}_{\bm a}(t-\tau).
 \ee
  Note that the system variable
 $W^{\sigma}_{a'}$ in \Eqs{delta_tilWa_mar}
 is now evaluated at the ending time of the path integral
 where $\alpha(t)=\alpha$ and $\alpha'(t)=\alpha'$ are fixed.
 As results, \Eq{delta_calR} for the Markovian-residue dissipation
 can be expressed in the operator level as
 \be\label{delcalRmar}
   \delta{\cal R}\, \hat O = \sum_{{\bm a},\sigma}
     [W^{\bar\sigma}_a,
     \delta{\bar C}^{(\sigma)}_{\bm a}\!(t)W^{\sigma}_{a'}\hat O
    - [\delta{\bar C}^{(\bar\sigma)}_{\bm a}\!(t)]^{\ast}
     \hat O W^{\sigma}_{a'}].
 \ee

 Alternatively, one can exploit various second-order QDT
 formalism\cite{Yan05187} for the residue
  $\delta R$ via, for example,
 \be\label{deltiW2}
   \delta{\ti W}^{(\sigma)}_{\bm a}(t)
  \approx
   \int_{t_0}^{t}\!d\tau\, \delta C^{(\sigma)}_{\bm a}(t-\tau)
  e^{-i{\cal L}(t-\tau)} W^{\sigma}_{a'}.
 \ee
 The dissipation-free propagator (assuming also
 time-independent system Hamiltonian for clarity)
 is used here to connect $W^{\sigma}_{a'}[\alpha(\tau)]$
 in \Eq{delta_tilWa} to its value at the fixed
 ending path point of $\alpha(t)=\alpha$.
 It is the reason that the $\alpha$-representation
 has been removed from \Eq{deltiW2}, as it is
 now treated as an operator; the same for
 $\delta{\ti W}^{\prime(\sigma)}_{\bm a}$.
 Therefore, the weak-residue $\delta{\cal R}$
 assumes the operator form as
 \be \label{calR2}
   \delta{\cal R}\,\hat O
  = \sum_{\bm a,\sigma} [W^{\bar\sigma}_a,
    \delta{\ti W}^{(\sigma)}_{\bm a}(t)\hat O
  - \hat O\, \delta{\ti W}^{\prime(\sigma)}_{\bm a}(t)].
 \ee
 Apparently, the Markovian-residue limit
 is the special case of the weak-residue treatment.

  To the end of this section, let us show that
 the principle of residue correction presented above
 can also be used for the hierarchy truncation.
 Recall the hierarchical AIF considered
 in the previous subsection, ${\cal F}_{\sf n}$
 of \Eq{AIFmar}. Implied there is a trivial identity that
 \be \label{FnplusFn}
   {\cal F}_{\check{\ind}^+_{(a),m}}\!\!
 = \check Z^{(a)}_m {\cal F}_{\sf n}
  = -i\big(\check W^{(a)}_m \pm \check W^{\prime (a)}_m\big)
  {\cal F}_{\sf n}.
 \ee
 Note that  $\check W^{(a)}_m$ and
 $\check W^{\prime (a)}_m$ defined in \Eqs{checkWm}
 as their path integral expressions are of the same
 mathematical structure as \Eq{delta_tilWa}.
 Therefore, the afore-described method, via either
 the Markovian or second-order approximation scheme
 to the explicit operator form of $\check W^{(a)}_m$,
 can be adopted {\it locally} at the desired anchoring
 level ${\sf N}$. Thus, the truncation scheme of
 \be\label{rhosfN_trum}
   \rho_{{\sf N}^+_{(a),m}}\!\! =\check Z^{(a)}_m\rho_{\sf N}=
  -i[\check W^{(a)}_m,\rho_{\sf N}]_{\pm},
 \ee
 in which
 $\check W^{(a)}_m$ assumes its approximate
 operator form is established.
 The choice of anchoring index-set ${\sf N}$
 will be specified at the concluding part of
 the coming section, where the hierarchical EOM formalism for
 the general non-Markovian dissipation is treated.

\section{Equations of motion theory for
   general non-Markovian dissipation}
\label{thgeneral}

\subsection{Influence generating functionals}
\label{thgeneralA}
  We are now in the position to the IGF--COPI construction
 of the exact hierarchical EOM formalism for the
 parameterized $C^{(\sigma)}_{\bm a}$ of \Eq{corr_para}.
 The corresponding dissipation functional can be expressed
 as [cf.\ \Eqs{calR_PI} and (\ref{calR_mar})]
 \bea \label{calRall}
  {\cal R}
  \!\!&=&\!\!
   i\sum_{\bm a,\sigma}
  {\cal W}^{\bar\sigma}_a Z^{(\bm a)}_{\D}
     + i\sum_{\bm a,\sigma,k} {\cal W}^{\bar\sigma}_a
  \big[X^{(\bm a)}_{k}+Y^{(\bm a)}_{k}\big]
 \nl \!\!&\ &\!
   +  i \sum_{a,\sigma,m}
     {\cal W}^{\bar\sigma}_a \check Z^{(a)}_{m}.
 \eea
 The last term arising from the Matsubara
 contribution had been treated in detail
 in \Sec{thspecialA},
 see \Eqs{calR_mar}--(\ref{checkcalCm}).
  Apparently, all these composite $\{X,Y,Z\}$-functionals
 are IGFs; but whether they are completed or not
 should be checked with their time derivatives;
 see \Sec{thspecialA}.

  Consider the Drude-IGFs in \Eq{calRall}, they are given by
 \bsube \label{ZD_PI}
 \be \label{ZDinPI}
   Z^{(\bm a)}_{\D}
  \equiv
   -i\left(\eta^{(\bm a)}_{\D}
   \ti W^{(\bm a)}_{\D}
      -\eta^{(\bar{\bm a})\ast}_{\D}
   \ti W^{\prime(\bm a)}_{\D}\right),
 \ee
 with
 \be \label{WD}
  \ti W^{(\bm a)}_{\D}
 =
   \int^t_{t_0}\!\!d\tau
    e^{-\gamma^{\bm a}_{\D}(t-\tau)}
   W^{\sigma}_{a'}[\alpha(\tau)].
 \ee
 \esube
 We obtain
 \be \label{dotZ}
  \partial_t Z^{(\bm a)}_{\D}
 = {\cal C}^{(\bm a)}_{\D} -\gamma^{\bm a}_{\D}
   Z^{(\bm a)}_{\D}.
 \ee
 Here
 \be\label{calCD_PI}
   {\cal C}^{(\bm a)}_{\D}
  \equiv
  -i\left(\eta^{(\bm a)}_{\D}W^{\sigma}_{a'}
  -\eta^{(\bar{\bm a})\ast}_{\D}W^{\prime\sigma}_{a'}\right).
 \ee
 Therefore, $\{Z^{(\bm a)}_{\D}\}$ constitute
 the complete IGFs as the Drude-term contribution
 is concerned.

  Turn now to the $\{X,Y\}$-IGFs. They are given by
\bsube \label{XY}
 \bea
    X^{(\bm a)}_{k}
 \!\!&\equiv&\!\!
    -i\big(\eta^{(\bm a)}_{2k}\ti W^{(\bm a)}_{2k}
      - \eta^{(\bar{\bm a})\ast}_{2k}\ti W^{\prime(\bm a)}_{2k}\big),
 \label{XinPI} \\
  Y^{(\bm a)}_{k}
 \!\!&\equiv&\!\!
    -i\big(\eta^{(\bm a)}_{2k-1}\ti W^{(\bm a)}_{2k-1}
   -\eta^{(\bar{\bm a})\ast}_{2k-1} \ti W^{\prime(\bm a)}_{2k-1}\big),
 \label{YinPI}
 \eea
\esube
 with [see \Eq{phik} for $\phi(t)$]
 \be \label{Waj}
  \ti W^{(\bm a)}_{j}
  = \int^t_{t_0}\!\!d\tau
     \phi^{\bm a}_{j}(t-\tau)
     W^{\sigma}_{a'}[\alpha(\tau)].
 \ee
 Unlike the $Z$-functionals for the Drude
 and Matsubara components,
 the time derivatives of the $X$- and $Y$-functionals are
 closed, together with two additional functionals, given by
 \bsube \label{XYbar}
  \bea
    \bar X^{(\bm a)}_{k}
 \!\!&\equiv&\!\!
   -i\big(\eta^{(\bm a)}_{2k}\ti W^{(\bm a)}_{2k-1}
  - \eta^{(\bar{\bm a})\ast}_{2k}\ti W^{\prime(\bm a)}_{2k-1}\big),
 \label{barX} \\
  \bar Y^{(\bm a)}_{k}
 \!\!&\equiv&\!\!
    -i\big(\eta^{(\bm a)}_{2k-1}\ti  W^{(\bm a)}_{2k}
   -\eta^{(\bar{\bm a})\ast}_{2k-1} \ti W^{\prime(\bm a)}_{2k}\big).
 \label{barY}
 \eea
\esube
 We have
 \bsube \label{dotXYZ}
 \bea
  \partial_t X^{(\bm a)}_{k}
  \!\!&=&\!\!
  {\cal A}^{(\bm a)}_{k} -\gamma^{\bm a}_{k} X^{(\bm a)}_{k}
   - \omega^{\bm a}_{k}\bar X^{(\bm a)}_{k},
\label{dotX1} \\
  \partial_t {\bar X}^{(\bm a)}_{k}
 \!\!&=&\!\!
    \omega^{\bm a}_{k}X^{(\bm a)}_{k}
   -\gamma^{\bm a}_{k}\bar X^{(\bm a)}_{k};
 \label{dotX2}
 \eea
and
\bea
 \partial_t {\bar Y}^{(\bm a)}_{k}
 \!\!&=&\!\!
 {\cal B}^{(\bm a)}_{k}
 -\gamma^{\bm a}_{k}{\bar Y}^{(\bm a)}_{k}
   - \omega^{\bm a}_{k}Y^{(\bm a)}_{k},
\label{dotY1} \\
  \partial_t Y^{(\bm a)}_{k}
 \!\!&=&\!\!
  \omega^{\bm a}_{k}{\bar Y}^{(\bm a)}_{k}
   -\gamma^{\bm a}_{k}Y^{(\bm a)}_{k}.
 \label{dotY2}
 \eea
 \esube
 Here
 \bsube\label{calABC_PI}
 \bea
  {\cal A}^{(\bm a)}_{k}
\!\!&\equiv&\!\!
  -i\left(\eta^{(\bm a)}_{2k}W^\sigma_{a'}
  -\eta^{(\bar{\bm a})\ast}_{2k}W^{\prime\sigma}_{a'}\right),
\label{calA_PI}
\\
  {\cal B}^{(\bm a)}_{k}
\!\!&\equiv&\!\!
  -i\left(\eta^{(\bm a)}_{2k-1}W^\sigma_{a'}
  - \eta^{(\bar{\bm a})\ast}_{2k-1}W^{\prime\sigma}_{a'}\right).
\label{calB_PI}
 \eea
\esube

  These six classes of $(X\!Y\!Z)$-functionals
 constitute now a complete
set of IGFs for the hierarchy construction.
 The general expression for all auxiliary influence functionals
 in the hierarchy can then be expressed as
 \bea\label{calFbl}
   {\cal F}_{\ind}
 \!\!&=&\!\!
  \Big\{\prod_{{\bm a},k}
  \Big[
       {\big(Y^{(\bm a)}_{k}\big)}^{n^{\!(\bm a)}_{2k-1}}
       {\big(X^{(\bm a)}_{k}\big)}^{n^{\!(\bm a)}_{2k}}
     {\big({\bar Y}^{(\bm a)}_{k}\big)}^{\bn^{\!(\bm a)}_{2k-1}}
     {\big({\bar X}^{(\bm a)}_{k}\big)}^{\bn^{\!(\bm a)}_{2k}}
  \Big]
\nl \!\!&\ &\, \times
  \prod_{{\bm a,\sigma}}
    \big(Z^{(\bm a)}_{\D}\big)^{n^{\!(\bm a)}_{\D}}
  \prod_{{a,\sigma,m}}
   \big(\check Z^{(a)}_{m}\big)^{\check n^{\!(a)}_{m}}
 \Big\} {\cal F}.
 \eea
The index in ${\cal F}_{\ind}$ is specified by the involving
 nonnegative integers,
 \be \label{indn}
  {\ind}\equiv
 \left(n^{\!(\bm a)}_{j}\!, \bn^{\!(\bm a)}_{j}\!,
  {n^{\!(\bm a)}_{\D}}\!, {\check n^{\!(a)}_{m}};
    \; {\mbox{\footnotesize{$j=1,\!\cdots\!,\!2K;
    \; m=1,\!\cdots\!,\!M$}}}
  \right).
 \ee
 Therefore, the total number of the nonnegative integers
 in the index-set $\ind$ is $2(p+4Kp + Mq)$,
 where $q$ denotes the number of dissipative modes,
 and $p\leq q^2$ the number of nonzero dissipative mode pairs.
  In terms of
 \Eqs{rhon_def} and (\ref{calUn_def}),
 the IGF--COPI approach to the hierarchical EOM is now
 completed by considering the time derivative
 of ${\cal F}_{\ind}$ [\Eq{calFbl}], which can
 be readily carried out using \Eqs{dotFinR},
 (\ref{dotcheckZm}), (\ref{dotZ}) and
 (\ref{dotXYZ}).

 \subsection{Hierarchical equations of motion}
\label{thgeneralB}

 The final hierarchical EOM for
 the associating auxiliary density operators read
\be \label{dotrhon}
 \dot\rho_{\ind}=-[i({\cal L}+\check n \mu)
   +{{\mathit\gamma}_{\ind}}]\rho_{\ind}
  +\rhonswap +\rhondown + \rhonup.
\ee
 The $\mu$-term in \Eq{dotrhon} arises from
 the corresponding contribution of Matsubara
 term [\Eq{dotcheckZm}] given by \Eq{checkn_mar}.
 The $\gamma$-terms in \Eq{dotrhon} arise
 from the damping-terms of \Eqs{dotXYZ}.
 The resulting damping constant is given by
 \bea\label{Gamind}
   {{\mathit\gamma}}_{\ind}
  \!\!&\equiv&\!\!
  \sum_{\bm a,k}
  \left[n^{(\bm a)}_{2k}+\bn^{(\bm a)}_{2k}
  +n^{(\bm a)}_{2k-1}+\bn^{(\bm a)}_{2k-1}\right]\gamma^{\bm a}_k
\nl \!\!& &\!\!
 +\sum_{\bm a,\sigma}
  n^{(\bm a)}_{\D}\gamma^{a}_{\D}
  +
 \sum_{a,\sigma,m}
 \check n^{(a)}_{m}\check{\gamma}_m .
 \eea

  The second term in \Eq{dotrhon}
stems from the (off-diagonal) swap-terms of
\Eqs{dotXYZ}. It reads
 \bea\label{rhonswap}
 \rhonswap
 \!\!&=&\!\!
 \sum_{\bm a,\sigma,k} \omega_{k}^{\bm a}
    \Big(
    n^{(\bm a)}_{2k-1} {\rho}_{{\ind}_{(\bm a),2k-1}^{\rightarrow}}
  - n^{(\bm a)}_{2k} {\rho}_{{\ind}_{(\bm a),2k}^{\rightarrow}}
 \nl && \qquad
  - \bn^{(\bm a)}_{2k-1}{\rho}_{{\ind}_{(\bm a),2k-1}^{\leftarrow}}
  + \bn^{(\bm a)}_{2k}  {\rho}_{{\ind}_{(\bm a),2k}^{\leftarrow}}
    \Big).
 \eea
The index-set ${\ind}^{\rightarrow}_{(\bm a),j}$
 and ${\ind}^{\leftarrow}_{(\bm a),j}$  differ from
 $\ind$ of \Eq{indn} only at the specified $\{(\bm a),j\}$ by
\begin{eqnarray*}
 {\ind}^{\rightarrow}_{(\bm a),j}: &&
  (n^{(\bm a)}_j,\bn^{(\bm a)}_j) \rightarrow
  (n^{(\bm a)}_{j}-1,\bn^{(\bm a)}_{j}+1),
\nl
 {\ind}^{\leftarrow}_{(\bm a),j}: &&
 (n^{(\bm a)}_{j}+1,\bn^{(\bm a)}_{j}-1)
  \leftarrow (n^{(\bm a)}_{j}, \bn^{(\bm a)}_{j}).
\end{eqnarray*}

 The third term in \Eq{dotrhon}
stems from the $(\cal{A,B,C})$-terms of \Eqs{dotXYZ}, while the last
term is from \Eq{dotFinR}. They are the hierarchy-down and
hierarchy-up contributions, respectively, and given by
\bea\label{rhondown}
  \rhondown
 \!\!&=&\!\!
 \sum_{\bm a,\sigma,k}\Big[
   n^{(\bm a)}_{2k}  {\cal A}^{(\bm a)}_{k}
       {\rho}_{{\ind}_{(\bm a),2k}^{-}}
  + \bn^{(\bm a)}_{2k-1}  {\cal B}^{(\bm a)}_{k}
       {\rho}_{{\bar{\ind}}_{({\bm a}),2k-1}^{-}}
    \Big]
 \nl \!\!&&\!\!
    +\sum_{\bm a,\sigma} n^{(\bm a)}_{\D} {\cal C}^{(\bm a)}_{\D}
       {\rho}_{{\ind}_{(\bm a),\D}^{-}}
 \! + \!
  \sum_{a,\sigma,m}\check n^{ (a)}_{m}\check{\cal C}^{ (a)}_{m}
  {\rho}_{\check{\ind}_{ (a),m}^{-}},
 \eea
 and ($j={\rm D},1,\cdots,2K$)
 \bea\label{rhonup}
 \rhonup \equiv -{\cal R}\rho_{\sf n}
 =  -i\! \sum_{\bm a,\sigma,j}\! {\cal W}^{\bar\sigma}_a
 {\rho}_{{\ind}^+_{(\bm a),j} }
   \! -i \!\! \sum_{a,\sigma,m}\! {\cal W}^{\bar\sigma}_a
    {\rho}_{\check{\ind}^+_{(a),m} }.
 \eea
  The reduced Liouville-space operator ${\cal W}^{\sigma}_a$
  in \Eq{rhonup} and $\check{\cal C}^{(a)}_m$ involved
 in \Eq{rhondown} were given by \Eq{calWcheckC}.
   In \Eq{rhondown}, ${\cal A}^{(\bm a)}_k$, ${\cal B}^{(\bm a)}_k$,
 and ${\cal C}^{(\bm a)}_{\D}$ denote
 the reduced Liouville-space operator counterparts
  of \Eqs{calCD_PI} and (\ref{calABC_PI}) 
 \bsube \label{calABC}
 \bea
  {\cal A}^{(\bm a)}_{k}\hat O
 \!\!&=&\!\!
   -i\big(\eta^{(\bm a)}_{2k}W^{\sigma}_{a'}\hat O
   - \eta^{(\bar{\bm a})\ast}_{2k}\,\hat O W^{\sigma}_{a'}\big) ,
 \\
  {\cal B}^{(\bm a)}_{k}\hat O
 \!\!&=&\!\!
   -i\big(\eta^{(\bm a)}_{2k-1}W^{\sigma}_{a'}\hat O -
   \eta^{(\bar{\bm a})\,\ast}_{2k-1}\,\hat OW^{\sigma}_{a'}\big) ,
 \\
 \label{CalCD}
 {\cal C}^{(\bm a)}_{\D}\hat O
 \!\!&\equiv&\!\!
  -i\left(\eta^{(\bm a)}_{\D}W^{\sigma}_{a'}\hat O
  -\eta^{(\bar{\bm a})\ast}_{\D}\hat O W^{\sigma}_{a'}\right),
 \eea
 \esube
 The index-set ${\ind}_{(\bm a),j}^\pm$ ($\bar{\ind}_{(\bm a),j}^{-}$
 or $\check{\ind}_{(a),m}^\pm$) differs from ${\ind}$ only by
 changing the specified $n^{(\bm a)}_j$ ($\bn^{(\bm a)}_j$ or $\check
 n^{(a)}_m$) to $n^{(\bm a)}_j \pm 1$ ($\bn^{(\bm a)}_j - 1$ or
  $\check n^{(a)}_m \pm 1$). Note that the
  $({\bar{\ind}_{(\bm a),j}^{+}})^{\rm th}$--auxiliary
  reduced density operators are not generated from
 \Eq{rhonup}, since $\bar X_k$ and $\bar Y_k$ do not appear in the
 influence phase integrand ${\cal R}$ [\Eq{dotFinR}]; they are rather
 generated from the EOM for the
  $({{\ind}_{(\bm a),j}^{+}})^{\rm th}$--auxiliary
 reduced density operators via the involved swap
 {\footnotesize\{${\leftrightarrows}$\}}--terms [cf.\ \Eq{rhonswap}].

  Note that, in \Eq{dotrhon},
  the reduced density matrix $\rho_0=\rho$ is of primary
 interest, and the initial conditions are
 $\rho_{{\sf n}}(t_0)=\rho(t_0) \delta_{\sf{n0}}$,
 as inferred from their definitions.
 Setting the initial time $t_0\rightarrow-\infty$,
 the established Hierarchical EOM is imposed no approximation.
 The initial conditions are now $\dot{\rho}(t_0)=0$,
 where $t_0$ can be at any time before applying the
 the external time-dependent fields. The pulse-field induced dynamics
 will then be evaluated via \Eq{dotrhon} in the presence of
 external field.

 \subsection{Comments}
\label{thgeneralC}
 \subsubsection{Residue dissipation}
    The hierarchical EOM, \Eqs{dotrhon}--(\ref{calABC}), are exact for
 a general dissipation system that involves the parameterized
 bath correlation functions of \Eqs{corr_para}.
 Due to the finite $K$ and $M$ of \Eqs{corr_para} in practice,
 the complete theory can be established by considering the residue dissipation
 $\delta{\cal R}(t)$, due to the small difference
 between the exact and the parameterized $C^{(\sigma)}_{\bm a}(t)$
  as described in section \ref{thresidue}.
 The final hierarchical
  EOM thus read [cf.\ \Eq{dotrhon0}]
 \be \label{dotrhonf}
 \dot\rho_{\ind}=-[i({\cal L}+\check n \mu)
   +{{\mathit\gamma}_{\ind}}+\delta {\cal R}(t)]\rho_{\ind}
  +\rhonswap +\rhondown + \rhonup.
 \ee
 Actually, this hierarchical EOM \Eq{dotrhonf}
 is valid for arbitrary temperature, including
  $T=0$, at which the Matsubara damping constant $\check{\gamma}_m$
 is meaningless and Matsubara expansion is not valid.
 Based on the principle of the residue correction,
 considering the finite temperature in the Matsubara term, the difference
 between the exact ($T=0$) and the parameterized
 $C^{(\sigma)}_{\bm a}(t)$ (finite $T$) can be incorporated
 in the residue dissipation $\delta{\cal R}(t)$.

 In the high temperature limit,
 a large Matsubara damping constant
 $\check{\gamma}_m\propto kT$ leads to the
 Markovian-limit bath correlation of $\check{C}^{\bm a}_m(t)$. In this case,
 the Matsubara-contribution can also be
 included in the residue dissipation $\delta{\cal R}(t)$ as
 given by \Eq{delcalRmar}
 which enter the theory globally at every hierarchical tier.

 \subsubsection{Truncation consideration}
 The hierarchy truncation via the principle of residue
 correction has also been demonstrated in subsection \ref{thresidue},
 where only the Matsubara term is considered.
 In addition to the truncation of the Matsubara term of \Eqs{FnplusFn} and
 (\ref{rhosfN_trum}),
 the same procedure is applied to the Drude
 and the $\phi$ terms
 ($\rho_{{\sf N}^{+}_{(\bm a),\D}} $ and $\rho_{{\sf N}^{+}_{{(\bm a)},j}}$).

 Now we consider the choice of the anchor index-set ${\sf N}$.
   The anchoring indexes can be specified  for the individual
  constituents of the interaction bath correlation functions
  $C^{(\sigma)}_{\bm a}(t)$ of \Eq{corr_para} as
\[
   N^{\bm a}_{\D},\,  N^{\bm a}_{k},\, {\check N}^{a}_{ m};
  \ \ \text{with\ } k=1,\cdots,K+1;  \ m=1,\cdots,M.
\]
 The closed set of hierarchically coupled EOM
 will then contain those $\rho_{\ind}$, whose
  individual index-set consists of the nonnegative integers
 that are confined within
 \bsube\label{confine_ind}
  \bea
   &\sum_{\sigma}n^{(\bm a)}_{2k}+n^{(\bm a)}_{2k+1}+
      \bar n^{(\bm a)}_{2k}
      +\bar n^{(\bm a)}_{2k+1} \leq N^{\bm a}_{k},
 \label{confine_indA} \\
  &\sum_{\sigma}n^{(\bm a)}_{\D} \leq N^{\bm a}_{\D}, \ \ \
   \sum_{\sigma}\check n^{(a)}_{m} \leq \check N^{a}_{ m}.
 \label{confine_indB}
  \eea
  \esube
 The anchor index-set ${\sf N}$ in $\rho_{\sf N}$
 can now be defined as those with
 at least one upper limit being reached.
 The truncation can thus be made based on
 the formally exact relations such as \Eqs{FnplusFn} and
 (\ref{rhosfN_trum}).

\section{Concluding remarks}
\label{thsum}

 In summary, we have generalized the
 exact non-Markovian quantum dissipation
 theory,\cite{Xu05041103,Xu07PRE} on the basis of the
 calculus-on-path-integral algorithm, to non-hermit
 coupling and non-Markovian grand canonical bath ensembles
 (both for Fermion and Boson statistics).
  The resulting dissipation
  functionals path-integral expression is
 found to be of the same as the canonical
 bath with the assumption of
 they all satisfying the Gaussian statistics.
 The distinct fluctuation
 dissipation properties lead to different
 hierarchical EOM coupling
 structures. However, the difference appears
 only in the Matsubara contributions, Boson-commutator versus the
 Fermion-anticommutator relations, due to the particularity
 of the Matsubara coefficients [\Eq{checketasym}].
  In the high temperature limit,
 the aforementioned distinction diminishes,
 as both the Fermion and Boson bath ensembles
 approach to the Boltzmann statistics.

%
%
%

  The hierarchical EOM formalism, which is in principle equivalent
 to its path-integral counterpart,
  may be relatively tractable; see the
 Summary of Ref.\ \onlinecite{Xu07PRE}.
  Its numerical implementation remains a challenge,
 especially for complicated, strong
 dissipation systems at the extremely low temperature regime.
  Note that the hierarchy is not
 just for system-bath coupling strength,
 but more importantly also for bath correlation
 time scale.\cite{Xu05041103}
 The proper choice of system to effectively reduce
 the non-Markovian system-bath
 coupling strength in \Eq{HT0}
 is therefore still important both physically and numerically here.
 The standard approaches in this regard include
 the variation-principle based canonical transformation
 (e.g.\ the polaron picture) and the primary-bath-mode method
 (e.g.\ the solvation coordinate picture).%
 \cite{Hol59325,Sil842615,Har851069,Pol9677,Goy05525}

 The present theory can account for the particle transfer between
 the system and the bath. Thus the problems of quantum transport through
 molecular wires and electron transfer in complex molecular systems
  can be demonstrated by this theory.
 For instance, if $W_a$ ($W^{\dg}_a$) represents annihilation (creation)
 operator of the system in the coupling Hamiltonian [\Eq{Hprmt}],
 the measurable quantity, such as the current can be exactly
 expressed as $ \la \hat I(t)\ra=-i\big\la[\hat N,H_{\rm T}(t)]\big\ra_{\T}
 =-i{\rm Tr}_{\s}\sum_{\bm a}
 \{W_a\rho^\dg_{\bm a}(t)-W^\dg_a\rho_{\bm a}(t)\}$.
 Here $\rho_{\bm a}(t)$ [$\rho^\dg_{\bm a}(t)$] is the
 summation over the
 reduced density matrixes in the first-tier described by
 \Eq{dotrhon} with $\ind=1$.
 This detail work will be treated elsewhere.
 The present theory can be further generalized to the system
 coupled with many reservoirs (including phonon bath).
 The work along this line is in progress and will be published
 elsewhere.


\begin{acknowledgments}
 Support from the NNSF of China
 (No.\ 50121202, No.\ 20403016 and No.\ 20533060),
 Ministry of Education of China (No.\ NCET-05-0546),
 and the RGC Hong Kong
 is acknowledged.
\end{acknowledgments}

\appendix
 \section{Quantum statistical properties
    of the interaction bath correlation functions: Derivation}
 \label{thapp_FDT}

 {\it The symmetry relation} -- the first identity of \Eq{Ctsym}.
 This can be obtained immediately from
 the definition of correlation functions,
 \Eq{ffcorr}, together with the time-translation invariance
 and the trace cyclic invariance.
 The details are as follows
 [noting that $\hat f^{\bar\sigma}_a \equiv (\hat f^{\sigma}_a)^{\dg}$].
 \bea \label{appFDT_sym}
  [C^{(\sigma)}_{ab}(t)]^{\ast}
  \!\!&\equiv&\!\!
   {\rm tr}_{\B}[e^{ih_{\B}t}\hat f^{\sigma}_a
    e^{-ih_{\B}t}\hat f^{\bar\sigma}_b\rho^{\rm eq}_{\B}]^{\ast}
 \nl  \!\!&=&\!\!
   {\rm tr}_{\B}[\rho^{\rm eq}_{\B}\hat f^{\sigma}_b
   e^{ih_{\B}t}\hat f^{\bar\sigma}_a e^{-ih_{\B}t} ]
 \nl  \!\!&=&\!\!
    {\rm tr}_{\B}[\hat f^{\sigma}_b(0)
     \hat f^{\bar\sigma}_a(t)\rho^{\rm eq}_{\B}]
 \nl  \!\!&=&\!\!
    C^{(\sigma)}_{ba}(-t).
 \eea

  {\it The detailed-balance relation} -- the second identity
   of \Eq{Ctsym}.
  Its derivation exploits the property of
  \be \label{fNmu}
   e^{\beta\mu\hat N}\hat f^{\sigma}_ae^{-\beta\mu\hat N}
  = e^{\sigma\beta\mu}\hat f^{\sigma}_a.
 \ee
 This relation arises from the linearity of $\hat f_{a}=\sum_j t_{aj}c_j$,
 together with the notions that $\hat f^{-}_a\equiv \hat f_a$
 and $\hat f^{+}_a \equiv \hat f^{\dg}_a$.

  Let us start with the derivation of \Eq{fNmu}.
 It is carried out by recognizing some basic
 quantum mechanics relations as follows.
 The first one is
 \be \label{ecalAB}
  e^{A}Be^{-A} = e^{\cal A}B = \sum_{n=0}^{\infty}
  \frac{1}{n!} {\cal A}^nB,
 \ee
 where ${\cal A}B \equiv [A,B]$. We have therefore
 \be \label{ecalNfsig}
   e^{\beta\mu\hat N}\hat f^{\sigma}_ae^{-\beta\mu\hat N}
  = \sum_{n=0}^{\infty}
  \frac{(\beta\mu)^n}{n!} \hat{\cal N}^n\hat f^{\sigma}_a,
 \ee
 where $\hat{\cal N}\hat f^{\sigma}_a\equiv[\hat N,f^{\sigma}_a]$.

  We shall also use the quantum mechanics relations of
 \be \label{abc_commutator}
   [AB,C] = A[B,C]_{\pm} \mp [A,C]_{\pm} B,
 \ee
 where $[\cdot,\cdot]_-\equiv [\cdot,\cdot]$ and
 $[\cdot,\cdot]_+\equiv \{\cdot,\cdot\}$
 denote the commutator and anticommutator, respectively.
 By noticing that $[c_k,c^{\dg}_j]_{\pm} =\delta_{kj}$ and
 $[c_k,c_j]_{\pm} =0$ for Fermions ($+$) or Bosons ($-$),
 we have $[\hat N, c_j]=\sum_{k}[c^\dg_ kc_k,c_j]=-c_j$
 and $[\hat N, c^\dg_j]=c^\dg_j$.
 Thus, $[\hat N,\hat f_a]=-\hat f_a$ and $[\hat N, \hat f^{\dg}_a]=
  \hat f^{\dg}_a$; i.e.,
 $ \hat {\cal N}\hat f^{\sigma}_a = [\hat N,\hat f^{\sigma}_a]
  = \sigma\hat f^{\sigma}_a$.
 It immediately leads \Eq{ecalNfsig} to \Eq{fNmu},
 which also implies
 \be\label{muNfbar}
  e^{\beta(h_{\B}-\mu\hat N)}\hat f^{\bar\sigma}_a(t)
  e^{-\beta(h_{\B}-\mu\hat N)}
  = e^{\sigma\beta\mu} \hat f^{\bar\sigma}_a(t-i\beta),
 \ee
 since $[h_{\B},\hat N]=0$ and $e^{\beta h_{\B}}f^{\sigma}_a(t)
   e^{-\beta h_{\B}} = f^{\sigma}_a(t-i\beta)$.

 The derivation of the detailed-balance relation,
 the second identity of \Eq{Ctsym}, is now straightforward,
 by using the last identity of \Eq{appFDT_sym}, the trace-cyclic invariance,
 \Eq{rhoeq} and \Eq{muNfbar}. We have
 \bea \label{appFDT_detbal}
   C^{(\sigma)}_{ba}(-t)
 \!\!&=&\!\!
   {\rm tr}_{\B}[\hat f^{\bar\sigma}_a(t)
     \rho^{\rm eq}_{\B}\hat f^{\sigma}_b(0)]
 \nl  \!\!&=&\!\!
    {\rm tr}_{\B}[\hat f^{\bar\sigma}_a(t)
     e^{-\beta(h_{\B}-\mu\hat N)}
    \hat f^{\sigma}_b(0)\rho^{\rm eq}_{\B}e^{\beta(h_{\B}-\mu\hat N)}]
 \nl  \!\!&=&\!\!
   \left\la e^{\beta(h_{\B}-\mu\hat N)}\hat f^{\bar\sigma}_a(t)
     e^{-\beta(h_{\B}-\mu\hat N)} \hat f^{\sigma}_b(0)\right\ra_{\B}
 \nl  \!\!&=&\!\!
     e^{\sigma\beta\mu}C^{\bar\sigma}_{ab}(t-i\beta).
 \eea
 The derivation of \Eqs{Ctsym},
 the symmetry and the detailed--balance
 relations in the time domain, is now completed.
 Their frequency-domain counterparts, \Eqs{Cwsym} with \Eq{Cwdef},
 are then followed immediately.

\section{Path integral formalism: Derivation}
\label{thapp_path}
  Let $U_{\rm T}(t,t_0;\{ f^{\sigma}_a(t)\})$ be
 the propagator of the total system and bath space,
 satisfying the \Sch equation of the total Hamiltonian
 of \Eq{HT0},
 \bea\label{app_Udt}
 \partial_t U_{\rm T}=-i\big[H(t)
  +\sum_{a,\sigma}W^{\bar\sigma}_{a}f^{\sigma}_{a}(t)\big]
  U_{\rm T}.
  \eea
 The total density operator at time $t$ is given by
 \be\label{rhos}
   \rho_{\rm T}(t)=
   U_{\rm T}(t,t_0;\{ f^{\sigma}_a(t)\})\rho_{\rm T}(t_0)
    U^{\dg}_{\rm T}(t,t_0;\{ f^{\sigma}_{a'}(t)\}).
 \ee
 Assuming it initially at time $t_0$ is [cf.\ \Eq{rhoeq}]
 \be \label{ansatz0}
   \rho_{\rm T}(t_0) = \rho(t_0) \rho^{\rm eq}_{\B}
   \propto \rho(t_0) e^{-\beta(h_{\B}-\mu\hat N)},
 \ee
 the reduced density matrix of primary interest is then
 \bea \label{rho0}
  \rho(t)
   \!\!&\equiv&\!\!{\rm tr}_{\B}[\rho_T(t)]
 \nl \!\!&=&\!\!
  {\rm tr}_{\B}
   \big[U_{\rm T}(t,t_0;\{f^{\sigma}_{a}(t)\})
   \rho(t_0) e^{-\beta (h_{\B}-\mu\hat  N)}
   \nl
   && U^{\dg}_{\rm T}(t,t_0;\{f^{\sigma}_{a'}(t)\})
   \rho^{\rm eq}_{\B}e^{\beta (h_{\B}-\mu \hat N)}
   \big]
 \nl
\!&=&\!
  \Big\la U_{\rm T}(t,t_0;\{e^{{\bar\sigma}\beta\mu}
  f^{\sigma}_{a}(t-i\beta)\})
     \rho(t_0) \nl
     &&U^\dg_{\rm T}(t,t_0;\{
     f^{\sigma}_{a'}(t)\}) \Big\ra_{\B}
 \nl \!&\equiv&\!
     {\cal U}(t,t_0)\rho(t_0).
\eea
 The third identity is obtained by using the
 trace cyclic invariance and the relations of
 \Eq{muNfbar}.

  Consider now the path integral expression of
 the reduced Liouville-space propagator ${\cal U}(t,t_0)$,
 as defined in the last identity of \Eq{rho0}.
 Let $\{|\alpha\ra\}$ be a basis set in the system subspace.
 We have
 \begin{eqnarray*}
   \!\!&\ &\!\!
  U_{\rm T}(\alpha,t;\alpha_0,t_0;\{f^{\sigma}_{a}(t)\})
 \nl \!\!&=&\!\!
  \int_{\alpha_0}^{\alpha}\!\!{\cal D}{\alpha}\,
   e^{iS[\alpha]} \exp_{+}\Bigl\{-i\sum_{a,\sigma}\!\int_{t_0}^{t}\!d\tau
    W^{\bar\sigma}_{a}[\alpha(\tau)]f^{\sigma}_{a}(\tau)\Bigr\} ,
 \nl
   \!\!&\ &\!\!
   U_{\rm T}^{\dg}(\alpha',t;\alpha'_0,t_0;\{f^{\sigma}_{a'}(t)\})
 \nl \!\!&=&\!\!
   \int_{\alpha'_0}^{\alpha'}\!\!{\cal D}{\alpha'}\,
   e^{-iS[\alpha']} \exp_{-}\Bigl\{i\sum_{\sigma,a'}\!\int_{t_0}^{t}\!d\tau
    W^{\bar\sigma}_{a}[\alpha'(\tau)]f^{\sigma}_{a'}(\tau)\Bigr\} .
 \end{eqnarray*}
 The action functional $S[\alpha]$ is related to the
 reduced system Hamiltonian $H(t)$ only.
 The system operators $\{W^{\sigma}_{a}\}$ have
 also been represented in path integral representation.
 On the other hand, the stochastic bath dynamics
 variables $\{f^{\sigma}_{a}\}$ remain in the operator form that
 involve in the time-ordered exponential functions.

 Substituting the above path integral expressions
 for \Eq{rho0} leads to \Eq{calGPI},
 with the influence functional there being given by
 \bea \label{app_FV}
    {\cal F}
 \!\!&=&\!\!
 \Big\la
      \exp_{+}\Bigl\{-i\sum_{a,\sigma}\!
    \int_{t_0}^t\!d\tau  W^{\bar\sigma}_{a}[\alpha(\tau)]
     e^{\bar\sigma\beta\mu}f^{\sigma}_{a}(\tau-i\beta)\Bigr\}
 \nl
     &&\times \exp_{-}\Bigl\{i\sum_{a',\sigma}\!
    \int_{t_0}^t\!d\tau  W^{\bar\sigma}_{a'}[\alpha'(\tau)]
    f^{\sigma}_{a'}(\tau)\Bigr\}
 \Big\ra_{\B}
 \nl \!\!&\equiv&\!\!
    \exp\left\{-\Phi[\alpha,\alpha']\right\} .
 \eea
 For $\{f^{\sigma}_{a}(t)\}$ satisfying Gaussian statistics, the
 bath ensemble average in \Eq{app_FV} can be evaluated exactly by
 using the second-order cumulant expansion method, as the higher
 order cumulants are all vanished.
 This property, together with \Eq{Ctsym},
 leads to the expression
 \bea\label{app_Phi}
 \!\!&\ &\!\!
 \Phi[\alpha,\alpha']
\nl \!\!&=&\!\!
 \sum_{a,a'} \int^t_{t_0}\!\!d\tau_2\!\!\int^{\tau_2}_{t_0}\!\!d\tau_1
 \Big\{W_a[\alpha(\tau_2)]
%
  W^{\dg}_{a'}[\alpha(\tau_1)]C^{(+)}_{a{a'}}(\tau_2-\tau_1)
 \nl && \quad
 + W^{\dg}_a[\alpha(\tau_2)]W_{a'}[\alpha(\tau_1)]
   C^{(-)}_{aa'}(\tau_2-\tau_1)
\nl && \quad 
  + W^{\dg}_{a'}[\alpha'(\tau_1)]W_a[\alpha'(\tau_2)]
    C^{(-){\ast}}_{aa'}(\tau_2-\tau_1)
\nl && \quad
  + W_{a'}[\alpha'(\tau_1)]W^{\dg}_a[\alpha'(\tau_2)]
  C^{(+)\ast}_{aa'}(\tau_2-\tau_1)\Big\}
\nl \!\!&&\!\! -
 \sum_{a,a'}\!\int^t_{t_0}\!\!d\tau_2\!\!\int^{t}_{t_0}\!\!d\tau_1
%
  \Big\{W_a[\alpha(\tau_2)]
   W^{\dg}_{a'}[\alpha'(\tau_1)]C^{(-)\ast}_{aa'}\!(\tau_2\!-\!\tau_1)
\nl && \quad
  +  W^{\dg}_a[\alpha(\tau_2)]W_{a'}[\alpha'(\tau_1)]
  C^{(+)\ast}_{aa'}(\tau_2-\tau_1)\Big\}
\nl \!\!&\equiv&\!\!
    \int_{t_0}^{t}\!d\tau {\cal R}[\tau;\{\bfalp\}].
 \eea
 for the influence functional exponent.

 In detail, we have used the symmetry relation
 [the first identity of \Eq{Ctsym}] in
 the third and fourth terms, and the
 detailed-balance relation [the second identity of \Eq{Ctsym}]
 in the last two terms of the above expression.
 Some elementary algebra will lead to \Eq{app_Phi},
 in terms of ${\cal R}\equiv \partial_t \Phi$, the expression,
 \bea\label{app_calR_PI}
 {\cal R}[t;\{\bfalp\}]
\!\!&=&\!\! \sum_{a,a'}\Bigl(
    \big\{W_a[\alpha(t)]-W_a[\alpha'(t)]\big\}
 \nl \!\!&\ & \ \times
  \big\{\ti W^{(+)}_{aa'}[t;\{\alpha\}]
      - \ti W^{\prime (+)}_{aa'}[t;\{\alpha'\}]
  \big\}
\nl &\ &\!\! +
  \big\{W^{\dg}_a[\alpha(\tau)]-W^{\dg}_a[\alpha'(\tau)]\big\}
 \nl \!\!&\ & \ \times
  \big\{\ti W^{(-)}_{aa'}[t;\{\alpha\}]
     -\ti W^{\prime (-)}_{aa'}[t;\{\alpha'\}]
  \big\}\Bigr)
\nl \!\!&=&\!\!
 \sum_{aa',\sigma}
    \big\{W^{\bar\sigma}_a[\alpha(t)]-W^{\bar\sigma}_a[\alpha'(t)]\big\}
 \nl \!\!&\ &\ \times
  \big\{\ti W^{(\sigma)}_{aa'}[t;\{\alpha\}]
      - \ti W^{\prime (\sigma)}_{aa'}[t;\{\alpha'\}]
  \big\},
 \eea
 with  $\ti W^{(\sigma)}_{aa'} \equiv \ti W^{(\sigma)}_{\bm a}$
 and $\ti W^{\prime (\sigma)}_{aa'} \equiv \ti W^{\prime (\sigma)}_{\bm a}$
 being given by \Eqs{tiW_PI}.
 Using the definitions of \Eqs{ticalW_PI} and (\ref{calW_PI}),
 the above expressions can be recast as \Eqs{calFfinal}.


\end{document}